# Primordial Planet Formation

Rudolph E. Schild[1,2]

[1]Center for Astrophysics, 60 Garden Street, Cambridge, MA 02138, USA

[2]rschild@cfa.harvard.edu

and

Carl H. Gibson[3,4]

[3]University of California San Diego, La Jolla, CA 92093-0411, USA

[4]cgibson@ucsd.edu, http://sdcc3.ucsd.edu/~ir118

**ABSTRACT**

Recent spacecraft observations exploring solar system properties impact standard paradigms of the formation of stars, planets and comets. We stress the unexpected cloud of microscopic dust resulting from the DEEP IMPACT mission, and the existence of molten nodules in STARDUST samples. And the theory of star formation does not explain the common occurrence of binary and multiple star systems in the standard gas fragmentation scenario. No current theory of planet formation can explain the iron core of the earth, under oceans of water.

These difficulties are avoided in a scenario where the planet mass objects form primordially and are today the baryonic dark matter. They have been detected in quasar microlensing and anomalous quasar radio brightening bursts. The primordial planets often concentrate together to form a star, with residual matter seen in pre-stellar accretion discs around the youngest stars.

These primordial planet mass bodies were formed of hydrogen-helium, aggregated in dense clumps of a trillion at the time of plasma neutralization 380,000 years after the big bang. Most have been frozen and invisible, but are now manifesting themselves in numerous ways as sensitive modern space telescopes become operational. Their key detection signature is their thermal emission spectrum, pegged at the 13.8 degrees Kelvin triple point of hydrogen, the baryonic dark matter (Staplefeldt et al. 1999).

**1. Introduction**

Recent discoveries of numerous planets seen orbiting ordinary stars in the local universe caught some, but not all, astronomers by surprise. The common view among professionals, shared with the general public, was that all such planets were formed in dust discs of the star as it formed and were caused by the star formation. Planets were seen orbiting in flat discs around the several stars with near-perfect edge-wise orientation to the human observer, many large and close to the star. An alternative view (Gibson, 1996) claimed all the hot, primordial, H-$^4$He gas emerging from the plasma epoch transition to gas would fragment under viscous-gravitational control to form planetary-mass gas clouds in dense clumps of a trillion. Rather than planets forming from stars, all stars should form from these primordial gas planets within their protostarcluster clumps.

It has been long recognized that most of the ordinary matter - protons, neutrons, and electrons as well as the atoms they constitute, are unseen and therefore were called baryonic dark matter. But although dark to most observations, the material had a gravitational signature from which it was understood to be present in large amounts in the Halo of our Milky Way Galaxy. So it was concluded that if it has a gravitational signature it should be possible to use the light bending properties of mass to detect the dark matter as gravitational lenses.

This idea received a strong boost when studies of the gravity associated with lensing galaxies seen along the paths to multiply imaged quasars gave a direct detection of the grainy texture of the lensing gravity, demonstrating that the mass of the lensing galaxy is dominated by objects with approximately the mass of the Earth, with stars making only a minor contribution (Schild, 1996). The mass could be estimated because the duration of events was quite short, approximately a day. The most extreme event found had an observed duration of only 12 hours, which, because of the cosmological expansion of time, corresponded to a time of only 5 hours referenced to the clock of an observer at the luminous quasar (Colley and Schild, 2003). Events caused by a microlensing star should have 30-year duration (Schild and Vakulik, 2005) and the frequently observed microlensing signal in Q0957+561 must be caused by a planetary object with a millionth of a solar mass.

Such rapid events represented a remarkable feat of nature. Recall that a quasar has 100 times the luminosity of an ordinary galaxy like our own Milky Way, and the observed 5-hour brightening and fading event seen as a 1% brightness change thus represents somebody effectively throwing a switch and all the luminosity equivalent to the billions of stars in a distant galaxy being switched on and off in 2 1/2 hours.

This is not really what happens, of course. It is a trick of nature operating through the Einstein General Theory of Relativity, whereby the gravitational field of a planet mass body comes perfectly into the line of sight to the distant quasar, and the gravitational field deflects the quasar brightness to a different direction, producing a brightness fading to some other observers at some other places. So although it was understood that the observation of a 5-hour microlensing event evidenced a multitude of free-roaming planets in the lensing galaxy at a redshift of 0.37, called rogue planets, they were in a cosmologically remote galaxy and not amenable to close study by direct observation. Rather than eight planets per star observed in the solar system the quasar microlensing result requires an average of many millions.

It was desirable to find such objects closer to home where they could be studied individually, and searches closer to home, in the Halo of our Galaxy where they had already been seen in aggregate stabilizing the rotation structure, but unfortunately most searches organized assumed a stellar mass, and the gravitational signature was therefore not found. (Alcock, C. et al. 1998). Even as the rapid brightness fluctuations corresponding to light bending from planet mass objects were being recognized in more lensed quasars, the Halo searches continued to disappoint. This was largely because it was being assumed that the dark matter would be uniformly distributed in space, contrary to the common observation that stars tend to cluster on all size scales, from double stars to star clusters to galaxies and their clusters.

The decisive observational clue came with a technological development, when detectors capable of measuring the feeble radiation of any object even slightly above that absolute zero of temperature were fitted to balloon-borne and orbiting telescopes. Already in 1985 the IRAS spacecraft had detected a feeble signature understood to be coming from all directions but unfortunately called "cirrus dust" clouds because of the way the radiation seemed to be a mottled distribution above and below the galaxy's disc (Low et al, 1984). Only when more recent generations of infrared and sub-millimeter detectors had been developed could the true nature of the feeble signature be understood.

This happened suddenly when maps of the sky taken by different spacecraft (COBE, IRAS, Boomerang, WMAP) operating in the same direction toward the Halo were intercompared (Veniziana et al. 2010) and the same structures seen. Because the different spacecraft detected different wavelengths, the inter-comparison allowed the temperature of the feeble structures to be measured, though at first still thought to be the infrared radiating dust clouds. Only when it was recognized that the temperature was being consistently found to be the same for all directions could the nature of the "infrared cirrus" be correctly inferred (Nieuwenhuizen et al. 2010). Figure 1 (Gibson 2010 Fig. 1) shows some of the Veniziani et al. (2010) evidence, compared to our interpretation that the "dust" of the "cirrus clouds" is actually primordial planets in their protoglobularstarcluster PGC clumps. The size of the observed clumps $4 \times 10^{17}$ meters closely matches the size of a globular star cluster.

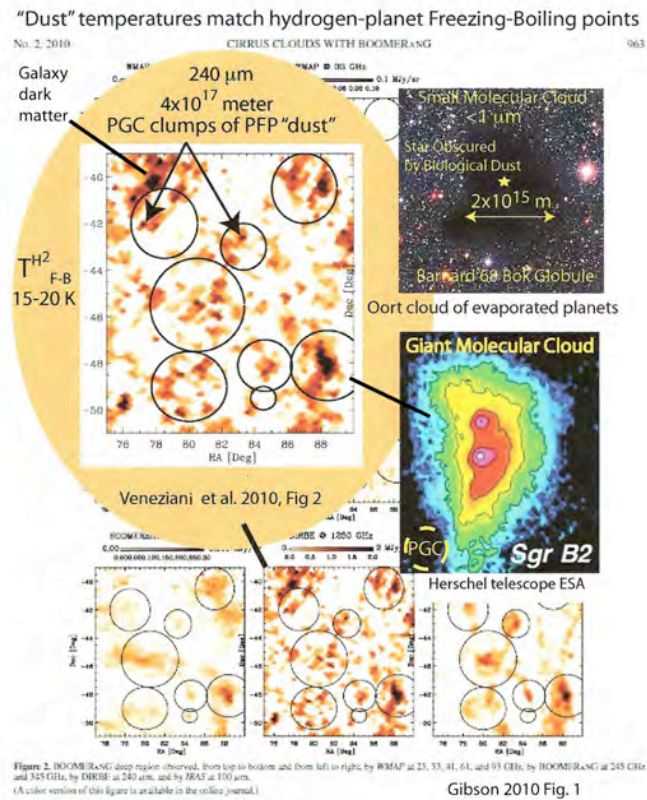

Fig. 1. Oval at left top is an image of so-called cirrus "dust" clouds of the Milky Way Galaxy (Veniziani et al. 2010, Fig. 2) at DIRBE microwave wavelength 240 μ compared to three other microwave images at the bottom showing the same features. The "dust" temperatures match the freezing to boiling point thermostat range of ∼ 15-20 K for hydrogen planets, therefore indicated as the "dust". Thus cirrus "dust" clouds are "Giant Molecular Clouds" (right center), presently interpreted as clumps of planet clump PGCs representing the Galaxy dark matter. The dark matter is obscured for optical wavelengths < 1 μ as shown (right top) for a star or stars obscured by an Oort cloud of microscopic PAH dust produced as planets are accreted and evaporated.

It is easy to see from Fig. 1 that the cirrus dust clouds are the Galaxy dark matter simply by estimating the number of PGC clumps of primordial planets from the images, realizing that each protoglobularstarcluster PGC clump represents a dark matter mass of more than a million stars. Comparing the DIRBE 240 μm image with the Herschel Sgr B2 image of a dense molecular cloud, we infer an average PGC density on the sky of about 20 PGCs per square degree. Since there are 38,800 square degrees on the sky, and since each PGC has mass ∼ $10^{36}$ kg, we estimate the "cirrus cloud" mass to be ∼ $10^{42}$ kg, as expected if the "dust" of the clouds is primordial planets and the "clouds" are PGC clumps.

Detection of µBD planetary objects from temperature happened because the temperature found was an important one in the physical chemistry of hydrogen, the most common element in the universe and the most probable candidate for the ordinary dark matter.  Hydrogen is a gas that condenses to liquid and then forms solid ice at low temperatures.   It has large latent heats of evaporation and fusion, so a thermostat temperature range between the triple point temperature and critical temperature is the signature of a cloud of gas planets cooled by outer space and heated by adjacent stars and friction.  A range of cloud temperatures between the triple point and critical point matches a range of planetary masses in dark matter planet clouds either cooled by outer space or heated by adjacent stars.  Thus if we imagine a primordial planetary atmosphere slowly cooling by radiating into space, it will remain at its triple point temperature for a long time as the hydrogen gas condenses to hydrogen snow and the heat of condensation slowly radiates away, producing a thermal signature pegged at the triple point temperature as observed.

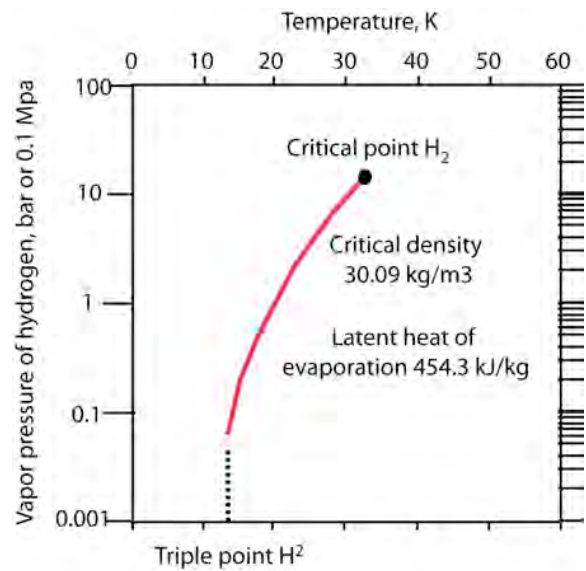

Fig. 2.  Vapor pressure of gas-liquid $H_2$ between its gas-liquid-solid triple point and gas-liquid critical point.  The large latent heat of $H_2$ evaporation thermostats PGC dark matter planet clouds to a 14-30 K triple point-critical point temperature range.

The hydrogen gas triple point temperature is 13.8 degrees Kelvin. The temperature inferred from the intercomparison of spacecraft observations (Veneziani et al, 2010) was inferred to be 15-20 K with small variation between the observed structures, about as expected. Thus the radiating mass in the halo of our galaxy can be inferred to be predominantly clouds of frozen hydrogen planets, particularly because nobody had ever predicted that "infrared cirrus" composed of interstellar dust grains would have this universally measured temperature. From Fig. 2, frozen hydrogen on microscopic dust grains in a vacuum would rapidly evaporate.

When this was realized, it was immediately obvious why temperatures measured for halos of other galaxies (going back to Price, Duric, & Duncan, 1994 with many similar determinations since) also were approximately 15 degrees Kelvin. Thus it is now apparent that a way exists to detect the ordinary dark matter anywhere it aggregates in the cold vastness of space, and this will provide an important tool to study the dark matter structures formed in galaxy collisions and in galaxy clumps, from its thermal emission. These temperature determinations are now common from PLANCK and HERSCHEL spacecraft data.

It is important to ask this question. If these hydrogen clouds are in fact the outer atmospheres of primordial planets that are the universal ordinary dark matter, at what time after the Big Bang (at what redshift) did they cool below the hydrogen triple point temperature? Recall that when the primordial planet clump structures seen today formed at the time when the universe had a temperature near 4000 K so that the expanding plasma had an important viscosity decrease 380,000 years after the big bang, freeing the gas to fragment into the planet-mass structures that would have been cooling ever since. Contraction would have re-heated their centers, whence monotonic cooling must have occurred, even as the background of remnant big Bang radiation would itself have been monotonically cooling, and is measured today to be 2.73 K. At some point the background temperature must have passed through the 13.8 degree hydrogen triple point, and the primordial planetary objects would have begun to freeze out hydrogen to become hydrogen ice which would have then snowed down through the atmospheres to the planetary surfaces. We will consider below how this produced dramatic weather effects on the primordial planets, but we wish here to discuss what the effects of the shrinking snowing atmospheres would have had on the UV-optical light transmission of the universe.

Given the present temperature of the universe and the rate the temperature falls with redshift, we easily calculate that the remnant Big Bang radiation and hence the temperature of the universe fell below the 13.8 degree hydrogen triple point temperature at redshift z = 6.0. It has already been noticed that distant quasars evidence a dramatic change in the transparency of the universe to hydrogen sensitive wavelengths at redshifts of approximately 6.3 (Fan et al, 2006). Presently this is attributed to a "re-ionization" of the hydrogen gas of the observable universe at this redshift, but with no explanation of exactly why at the particular observed redshift of 6.3. In particular, there had been no prediction that the transmission of the universe would have dramatically increased at z = 6.3. The possibility that the transparency of the universe should be significantly affected by the atmospheres of a population of primordial planet clouds has already been considered by Schild and Dekker (2006).

We predict that if the hydrogen transparency increase of the universe is due to shrinking primordial planet atmospheres, other observational manifestations should be discovered. For example, the shrunken planets would have lower cross sections for collisions with implications for changes in their collisional dynamics, increasing diffusion of the PGC planet clouds from the galaxy central core to form the galaxy dark matter halo. With suddenly smaller atmospheres there would be fewer planet collisions and Ofek et al. (2010) events as the star formation rate decreases. In some ways it can be seen that the cooling of the universe below the hydrogen triple point represents a phase change, which will have implications for the rate of aggregation into stars and in their collisional interaction signatures (Nieuwenhuizen et al. 2010).

## 2. Detection of Primordial Planets

As the population of primordial planets has been evolving throughout the history of the universe, their detection will be complicated according to the redshift being probed. In the local universe, we see the objects in experiments related to microlensing of quasars, where a distant quasar at a redshift of approximately 2 is seen aligned with a foreground galaxy at a typical redshift below 1. As the global gravitational field of the lensing foreground galaxy creates multiple images of the distant quasar, the individual objects within the lensing galaxy in line with the background quasar can modify the quasar light further. This created signatures of any mass bodies in the lensing galaxy, and signatures of stars (Schild & Smith, 1991). Planets (Schild 1996), and star clusters (Mao and Schneider, 1998) have been recognized.

Here we are particularly interested in the planet signatures. Because of the small brief structures caused by quasar-planetary alignments, the planet signatures are discussed under the topic of microlensing, or even nanolensing, and the first demonstration of the phenomenon operating at planetary scale by Schild (1996) in quasar Q0957+561 was quickly followed by discoveries in other gravitational lens quasar systems. In Q2237 (Vakulik et al. 2007), a microlensing event from a 0.001 $M_{Sun}$ free-roaming (Jupiter mass) planet was found. Systems 0B1600+43, (90 day microlensing, 0.0001 $M_{Sun}$ (Burud et al, 2000 ), RXJ 0911.4, 90 days, 0.0001 $M_{Sun}$ (Hjorth et al, 2002), FBQ 0951 (Jakobsson, et al, 2005, SBS1520 and FBQ 0951 (90 day, 0.0001 to 0.0001 $M_{Sun}$ (Paraficz et al, 2006) produced similar detections.

By far the best brightness monitoring data revealing the microlensing signature is from the first discovered and longest monitored system, Q0957+561. This is because it has the best-determined time delay, accurate to 0.1 day (Colley et al, 2003). Recall that the quasar microlensing is revealed by comparing two images of a multiply imaged quasar and correcting for the delay between arrival of the multiple images. The only quasar with delay measured to such 0.1 day precision is Q0957+561, following an intense monitoring effort by an international consortium (Colley et al. 2003). This allowed Colley and Schild (2003) to securely detect an event with a 12-hour total duration, the most rapid ever detected. A wavelet analysis of the longer brightness record in this quasar lens system demonstrated several hundred microlensing events (Schild 1999).

But because these were all planet mass bodies seen in remote galaxies, it was desirable to detect the objects in the Halo of our own Galaxy. Unfortunately the results of 2 research consortia that sought the primordial planet signal failed in spite of monitoring 7 million stars in the Large Magellanic Cloud for the microlensing signature. While published accounts make the claim that the objects do not exist and ignore the many detections in the quasar lens systems, it is easy to understand that the local failures result from systematic errors of assumptions, like the assumption that the hydrogen atmosphere would be too small to cause refraction, or the assumptions about the sizes of the stars monitored. Microlensing consortia (MACHO, EROS, OGLE etc.) have made the fatal assumption that the planetary objects they claim to exclude are uniformly distributed: not clumped and clustered as observed in Fig. 1. Thus, they have excluded nothing. New observations are planned focusing on stars with foreground planet clouds.

## 3. Weather on Heating and Cooling Primordial Planets

Although halo searches could not find the free-roaming planet mass bodies, many have since been found in star formation regions in the vicinity of the sun (Lucas et al, 2006; Marsh et al, 2010, Zapatero Osario, 2002). They would only be found in young star-forming regions because after their compressional heating they quickly cool and become no longer visible. Their present properties of temperature, spectrum, color, and luminosity have been measured and masses inferred to be planetary from model fitting. Objects so recently formed have normal solar metal compositions, unlike the primordial objects in general.

Many of the weather effects expected on the primordial planets would be related to weather phenomena known on earth and on other solar system planets. On Earth, latent heats of fusion and evaporization of water rather than hydrogen are important weather drivers, causing local heating of gas masses (clouds) and strong vertical motions. Freezing point temperatures of water ice are clear in clouds where rain and hail are formed. Horizontal winds are produced when atmospheric gases replace the vertical convective elements. The convecting elements are internally heated by latent heat of fusion and vaporization, and become driven out of local thermal equilibrium, further driving weather phenomena.

Images of primordial planet atmospheres formed by star heating are available from space telescopes. Figure 3 shows images of primordial planets evaporated near a white dwarf star in our nearest planetary nebula Helix.

Oort cavities are formed within PGC clumps of planets by the star formation process of accreting planets. As shown in Fig. 3, the Oort cavity size is $(M_{star}/\rho_{PGC})^{1/3} \sim 3 \times 10^{15}$ m, where $M_{star} \sim 1.5 \times 10^{30}$ kg is the mass of the star formed and $\rho_{PGC}$ is the density of PGC planet clumps $\sim 4 \times 10^{-17}$ kg/m$^3$. Biological PAH dust from the evaporated planets fills the cavity, as seen by the Spitzer infrared space telescope image (bottom right). Smaller solar system scales are shown for comparison (bottom left). Matsuura et al. (2009) show hydrogen resonant images that reveal primordial planets and their wakes as well as planet proto-comets on their way to accretion that are invisible in the optical band Hubble space telescope image. The planets ejected by the plasma jet must refreeze when they reach distances of $\sim 10^{16}$ m from the central star and begin radiation to outer space.

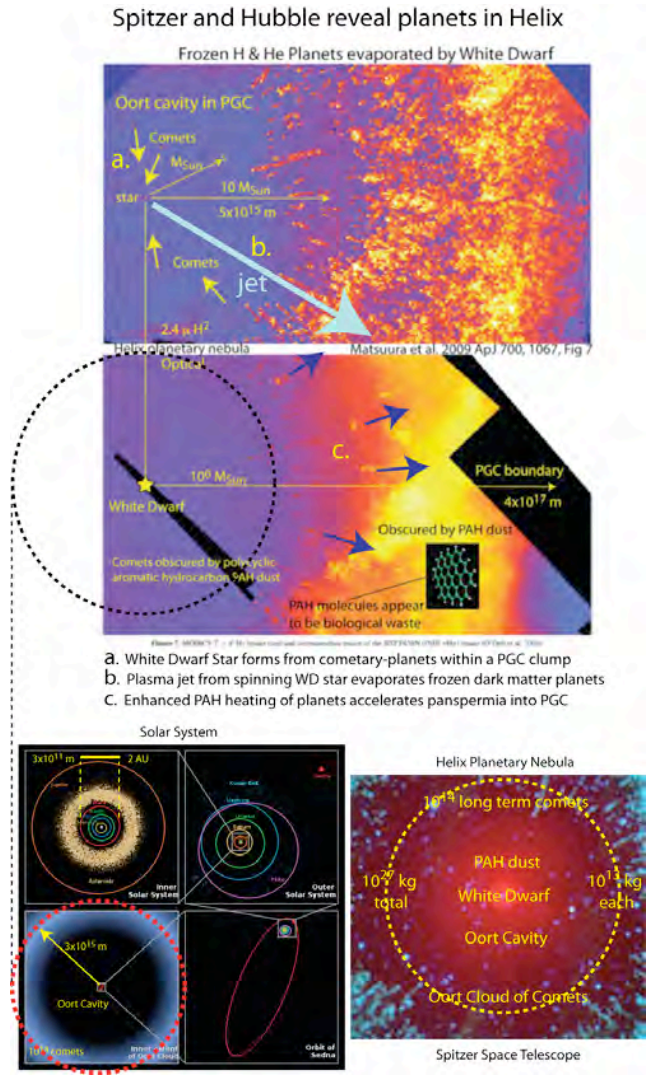

Fig. 3. Primordial planets heated and cooled in nearby planetary nebula Helix. A central white dwarf star is overfed by accreting primordial planets so that it contracts to high density $\sim 10^{10}$ kg/m$^3$ and its spin rate increases. A powerful plasma jet (blue arrow top) is produced that evaporates and radially expels some planets bounding the Oort Cavity. Some continue to fall into the cavity as comets (top image for a 2.4 μ infrared H$_2$ resonance).

New classes of weather are expected for the heated and cooled primordial planets of Fig. 3. Terrestrial weather is largely driven by direct heating of the Earth atmosphere and surface due to the absorption of sunlight, and the earth core is being heated from tidal forces. The primordial planets, in contrast, monotonically cool over the history of the universe, with external heating when stars are formed nearby. Furthermore the hydrogen atmospheres will initially extend far beyond the solid cores, where the Earth has lost most of its atmospheric gases, particularly hydrogen, and the present earth atmosphere extends only a small fraction above its solid-core radius.

This will create the potential for much larger velocities and more violent atmospheric effects. We recall that the escape velocity on earth is 11 km/sec. Thus infalling meteors and meteorites will have this velocity plus an "inter-solar system orbital velocity" of order 50 km/sec. Dissipation of the kinetic energy of these interlopers will cause local heating effects in the atmosphere and produce locally high winds.

Terrestrial falling snow is limited to terminal velocity speed dominated by atmospheric friction. For the primordial planets there would presumably be a strong radial center-to-edge temperature and density gradient, and hydrogen ice would ordinarily form at the top of the large atmosphere. Falling hydrogen ice should accumulate and create snowballs of unknown size and density contrast over their long journey to the surface, perhaps even becoming crystalline and comparable to hail stones, with potential for larger velocities limited to the escape velocity, but perhaps limited to the local sound speed.

Recall that observations of solar system comets seem to show surfaces of compressed icy solids and powdery dust (Kearsley et al. 2008). Mean densities of comets are approximately a quarter of terrestrial crystalline rock densities, suggesting a powdery consistency throughout. This suggests that the surfaces of Oort Cloud comets were formed at the time of formation of the sun, in hydrogen ice snowstorms.

Modern observations of Kuiper belt and Oort cloud objects allow sizes to be measured, and mean densities to be inferred in cases of binary objects. With the understanding that these outer solar system objects had a primordial origin, where a contracting hydrogen cloud developed a heated center that was hotter for larger objects, it is possible to make a prediction about measured densities as a function of total mass. In this picture the most massive objects had the hottest core and the largest hydrogen envelope, so they remained longest with elevated temperatures and attracted the most dust which would melt at the central core. Subsequent slow evaporation of hydrogen would have left relatively more time to add to the central core of molten stratified dust. And since the present day cool object has a relatively larger fraction of metals in solidified crystalline material, the heavy remnants will have a higher mean density than the lighter remnants, which accumulated more dust on the surface where it remained a more porous dust layer.

These speculations are only intended to remind that strong weather forces would be expected where the primordial planets have potentially larger atmospheres and a potentially large overall temperature gradient. With strong forces potentially in play, stratified turbulence is also likely, as will be discussed in a companion report focused upon the analysis of atmospheric forces and universal similarities. A variety of stratified turbulence and turbulent mixing signatures within the planet cluster containing the sun have been identified from radio telescope signals (Gibson 2010).

## 4. When and how did planets form?

The traditional view of the formation of the Earth and all planets seen orbiting stars is that formation began in pre-stellar discs around newly forming stars. The only ingredients presumed available are microscopic interstellar dust particles, plus primordial gases dominated by hydrogen. This is supposed to occur within "cores" of large gas clouds having sufficient mass to make hundreds or thousands of stars.

However there is presently no theory of where the gas clouds came from in the first place, and the process by which cores form is not understood, since it is believed that the predominantly hydrogen gas cannot cool rapidly enough to prevent compressional heating generating pressure that would re-expand the cloud core.

Recent developments in understanding of hydrodynamical forces on the expanding Big Bang gas with dissipative viscous forces produce a very different picture (Gibson, C. & Schild, R. 2003). Within low $10^{-5}$ density enhancements structure formed by the condensation-void separation process on mass scales of galaxies to galaxy clusters before recombination. At plasma neutralization, 380,000 years after the Big Bang, the entire baryonic matter content of the universe underwent a very large decrease in viscosity, freeing up the gas to further condense on mass scales of planets (primordial-fog-particles, PFPs, in the discovery reference Gibson 1996) and of cluster mass aggregates with $10^6$ $M_{Sun}$. Notice that this produces a hierarchical clumping of mass on scales of planets within clusters within galaxies within clusters of galaxies. But in particular, no stars formed primordially within this framework.

In this picture, the newly formed structures in quiescent universal expansion would develop small random motions, and as the process of separation and condensation was underway locally, some of the cluster scale perturbations would have interacted, jostling the contained primordial planets and causing some to collide and stick together, initiating pairing of pairs of pairs etc. for a bottoms-up aggregation that resulted finally in a structure massive enough that central temperatures climbed above the fusion temperature and density, and a star began to shine. Statistics describing the ubiquity of presently observed orbiting double/multiple stars and binary Kuiper belt objects result from this pairwise aggregating process. The ubiquity of dynamically double and multiple gravitating systems (Kern & Elliot, 2006) defy the simplistic gas fragmentation hypothesis of star and planet formation.

At the time of formation, 380,000 years after the big bang, the primordial planets would have been predominantly hydrogen gas spheres, with a radius comparable to the separation distance of $10^{14}$ m. Since then they would have gradually cooled and shrunk. Where these objects are seen near white dwarfs in planetary nebulae of the local universe like the central star of Helix (Fig. 3) they are called "cometary knots". The planet atmospheres are of size $10^{13}$ m (Meaburn et al, 1992), which is 100 A.U. Recall that a million of these had to aggregate to form the sun, which is compatible with the known diameter and volume of the Oort cloud. Such large objects would have swept up the interstellar medium of the Galaxy disc and Halo (Schild 2007) so the initial hydrogen composition would evolve to higher metal richness in parallel with the stellar population of the Galaxy.

The collection of interstellar dust by this mechanism can solve several problems in the planet formation scenario. Recall that the standard picture requires dust grains to collide and stick together, whence the larger grains also collide and eventually aggregate as planet mass bodies. The problem is that this process of colliding and sticking must be highly efficient for all mass scales, dust to pebbles to boulders to mountains to planets, since if it fails for any mass scale, the entire process breaks down. But colliding rocks together always produces simple recoil or fracturing. In the history of our civilization, nobody has reported two rocks colliding together and sticking. This point will be further discussed in the following section.

## 5. Iron cores of Planets

The theory of the formation of the planets is confounded by two observational facts. The widely accepted theory considers that planets were formed in pre-stellar discs that can be seen as the equatorial plane of several nearby stars with favorable orientation. Such discs can be found from infrared signatures in young star forming regions, and are known to dissipate and disappear on time scales of 5 million years (Currie et al. 2010).

However the mechanism by which the planets form within the disc is problematic. It is postulated that dust grains of oxides of iron and silicon etc found in interstellar space collide and stick together, producing larger grains that collide and stick, in an accretional cascade that eventually produces planets.

There is a serious problem with this scenario. The experiment has been done billions of times in our civilization, and rocks colliding with rocks have never been observed to stick together; they always recoil or break up. But the process must work efficiently for all sizes of colliding rocks from interstellar dust grains to stones to rocks and mountains, or the process breaks down. Such an accretional cascade seems impossible.

Attempts to create the process in a laboratory have failed, as expected. These experiments have been summarized by Blum and Wurm (2008). It takes a careful reading to understand that the "Porosity index" quoted, consistently shows that the marginally sticking fragments are not held together by crystalline forces, but rather by much weaker van der Waals forces, and in a reference cited as "Schraepler and Wurm, unpublished data" it is noted that a single ballistic inbound particle breaks up the weakly bound chains and clumps, and the process breaks down.

This problem is sidestepped if the planets are formed in a two-step process, with the first step occurring after plasma neutralization and the second in the pre-stellar accretion discs around young forming stars. In the first step, which occurred in the long history following plasma neutralization at $z = 1000$ until approximately $z = 1$ when the sun formed, the primordial rogue planets roamed the Halo of the Galaxy and their large hydrogen atmospheres swept up all the interstellar dust distributed by supernovae and by carbon stars (Schild, 2007). The dust grains meteored into the hydrogen atmospheres and collected at the centers because of their metallic densities. Later, in the prestellar disc phase, strong interactions caused them to shed their hydrogen atmospheres to create the sun, and further collisions among them modified the population further, including removal of the residual hydrogen by evaporation.

A second important problem of planet formation is the iron core of the Earth. Anyone seeing rusting hulks of junked automobiles knows that iron has a strong propensity to oxidize, and every steel-maker knows how difficult it is to reduce iron oxide to base metal; it is done in high temperature blast furnaces in the presence of hot reducing carbon monoxide.

The metallic ions in interstellar dust are always found as oxides, so when and where in the universe was this processed to become iron cores of planets surmounted by oceans of water? This would naturally have happened when the mBD population of planet mass objects with very large hydrogen atmospheres roamed the universe and swept clean the interstellar medium. Interstellar dust grains would have meteored into the extended cooling hydrogen atmospheres and vaporized. The hydrogen is known to reduce any metal to form water which precipitates out, and the dense metallic vapors settled to the mBD center to stratify with all other metallic ions according to density, beneath an ocean of water.

These events would have been most common in the early universe, before $z = 6$ when cooling of the universe below the hydrogen triple point caused the atmospheres to freeze out hydrogen as snow and reduce the sizes of the atmospheres. At higher redshifts, $6 < z < 1000$, the planets would have had frequent collisions and mergers in their accretional cascades that sometimes produced stars. Release of gravitational binding energy would have re-heated the primordial planets, particularly their metallic cores, and the initially stratified reduced metals would have been smeared somewhat to become the ore deposits mined today.

Over the duration of the universe, the nature of planets would have evolved. At initial formation, such primordial planets would have had large sticky atmospheres but negligible relative motion in a quiescent expanding universe. This would have resulted in some collisional accretion even before the universe had created metals, and locally violent interactions would have produced further accumulations, generating a mass spectrum with a long tail toward higher sub-stellar masses but ultimately also stellar masses.

In this way, the first generation Pop III stars formed gently and early from hot large atmosphere gas planets of almost pure hydrogen. Their first generation of supernovae, plus the collapsing galaxy-central objects soon to become black holes after passing through a temperature realm sufficiently high to fuse hydrogen, would have been the first luminosity in the universe, which at z = 1000 had cooled below 4000 K. With further expansion and cooling of the Universe, but also with accumulating larger relative motions, interactions of the primordial planets and their stars gradually increased.

In the hydrogravitational dynamics HGD picture of global structure formation, the above local scenario of star formation would have increasingly been affected by the events that shaped the formation and structures of galaxies. Over time, such events would have impressed larger relative velocities onto primordial planets and stars, increasing interactions and further accelerating accretional cascade from planets to stars. In addition, matter was also clumping from a redshift 1000 onwards on scales of Jeans Clusters, and just as the initial interactions of the quiescent expansion were few, there would have been some at first. The result would have been that a few such cluster mass objects would have interacted and the violent velocities generated would have triggered extensive star formation in globular clusters. With matter initially clumped as mBDs in Jeans Clusters, the initial interaction of two clusters would have strongly increased the process of collisional cascade, and the first luminosity of the universe would have been dominated by these collisionally interacting proto-globular-star-clusters. This would explain why the primordial globular clusters have profoundly different star formation histories and dynamics than the more recently forming open clusters. Without the primordial formation picture of Jeans mass structures, the origin of globular clusters has never been understood.

Subsequent star formation in the universe would have been predominantly in the Jeans cluster mass clouds fragmented into mBDs and their primordial sub-stellar cascade products, which are principally objects of less than Jupiter mass. This is inferred from the fact that the recently discovered planets found orbiting stars in the solar vicinity have a sharp upper mass limit of $\sim 10^{28}$ kg $\sim 18$ $M_{Jupiter}$ (Watson et al, 2010). But without the initial strong interactions of the HGD structure formation picture, the difference between the presently forming open clusters and the ancient globulars has never been explained.

In the present day star formation in open clusters, stars are seen commonly forming as double and multiple stars, where the gas fragmentation picture does not generate such multiple systems. In the HGD scenario with primordial mBD formation, the multiple star seeds formed long before a recent jolt triggered star formation.

## 6. The HGD universal spectrum of mass fragmentation

A surprising result of the hydrogravitational dynamics HGD scenario of structure formation is the way the universe develops a fragmenting hierarchy of structure, with its key elements arranged in a top down sequence of structures of decreasing mass between voids expanding at the plasma sonic velocity near light speeds. It reflects the common observation that galaxies are usually found in clusters of galaxies, globular star clusters are in galaxies, and stars are found in clusters. It is clear that this should be true for the HGD fragmentational schenario since the instability of density minima is enhanced by the expansion of space but condensations on maxima is inhibited. The opposite is expected for the standard (but false) ΛCDM scenario where gravitational structures build up from small masses to large controlled by (impossible) cold dark matter CDM condensations and superclusters are the last rather than first structures to form.

We show in Figure 4 the predicted viscous fragmentation mass scales of the plasma and gas epochs (Gibson 1996). The length scale of fragmentation is determined by a balance between viscous and gravitational forces at the Schwarz viscous scale $L_{SV} = (\nu\gamma/\rho G)^{1/2}$, where $\nu$ is the kinematic viscosity, $\gamma$ is the rate-of-strain of the fluid, $\rho$ is the density and G is Newton's gravitational constant. For the first plasma gravitational structures to form, $L_{SV}$ must become smaller than the scale of causal connection $L_H = ct$, where $c$ is the speed of light and $t$ is the time since the cosmological big bang. This occurs at time $t \sim 10^{12}$ seconds (about 30,000 years).

Momentum is transported in the plasma by photons that collide with free electrons, that in turn drag along their associated ions (protons and alph-particles). The kinematic viscosity ν is estimated (Gibson 2000) to be $\sim 10^{26}$ m² s⁻¹, γ is $\sim 10^{-12}$ s⁻¹ and ρ $\sim 4 \times 10^{-17}$ kg m⁻³. so $L_{SV} \sim 2 \times 10^{20}$ m $\sim L_H = 3 \times 10^{20}$ m. The mass scale is $\sim 10^{45}$ kg; that is, the mass of a supercluster of $\sim 10^3$ galaxies. The density ρ remains constant in the fragments with time but the masses of the subsequent fragments decrease because both ν and γ decrease as the universe cools and expands. The smallest plasma fragments are galaxies as shown in Fig. 4 on the right.

Notice, however, that the HGD structure formation scenario also has important consequences for the microlensing detection of its component objects. A concentric hierarchical structure boosts and increase on several scales at the same time. Insofar as the studies appear smaller than their projected Einstein rings, each consecutive structure increases magnification. The compounding of images behind a network of rogue planets behind a microlensing star has already been simulated by Schild and Vakulik (2003), but in principle the further lensing by a star cluster and galaxy need to be considered.

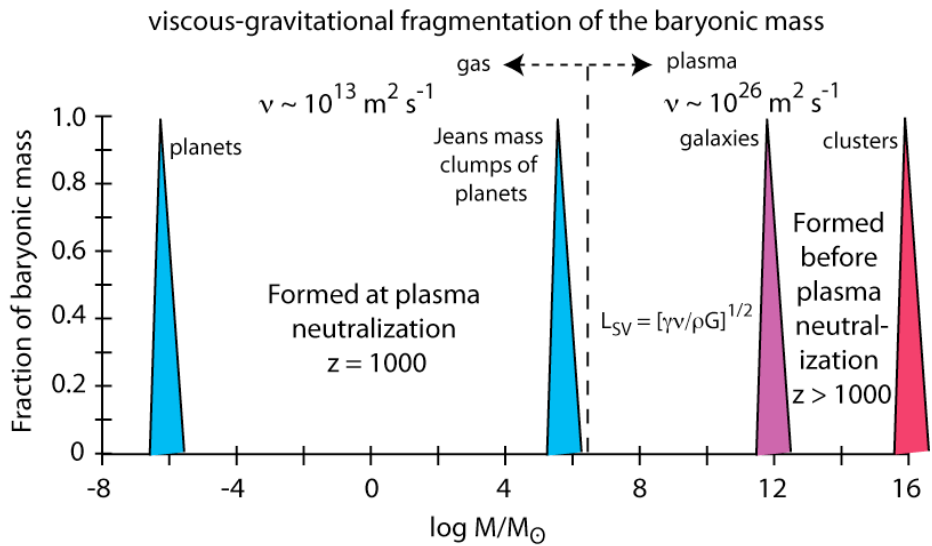

Fig. 4. Fragmentation of the baryonic mass under viscous-gravitational control. The first structures to fragment were at super-galaxy-cluster mass, when the horizon scale of causal connection ct exceeded the viscous-gravitational scale $L_{SV}$ in the plasma epoch (right). At later times in the plasma epoch, lower mass concentrations down to galaxy

> mass would form (Gibson 1996). A dramatic decrease in kinematic viscosity at plasma neutralization accounts for the decrease in fragmentation mass from that of galaxies to that of planets in clusters.

In the plasma epoch, galaxy super-clusters formed first, and subsequent structure formation decreased at plasma neutralization to a minimum mass scale of galaxies ($10^{42}$ kg) within clusters, and with characteristic $\delta T/T$ amplitude of $10^{-5}$. With the severe viscosity drop to $\nu \sim 10^{13}$ m$^2$ s$^{-1}$ at the time of plasma neutralization $10^{13}$ seconds, internal structure formed on scales of Jeans clusters and planets, but not stars. The latter would be formed in an accretional cascade of planet mass hydrogen spheres, as mathematically simulated by Truelove (1997) who, however, treated them as "numerical instabilities" in his calculation. With a turbulent hydrodynamical formation scenario, it would be unsurprising if the scales of structure formation in the universe, from galaxy super-clusters, galaxies, star clusters, to planets, follow a self-similar mathematics (Oldershaw, 2010).

The remainder of this section describes how to reasonably understand the broad range of Solar System objects found in the local universe: planets, KBO's, asteroids, comets, and meteorites.

At the time of plasma neutralization, the temperature of the universe was ~ 4000 K and no metals existed. As the universe cooled, a first generation of massive stars resulted in supernova enhancement of the interstellar medium, and the hydrogen gradually became contaminated with metals. This would have happened on a time scale of 10 million years (Gibson, 1996; Gibson and Schild, 2010).

From the known objects studied within our Solar System, we infer that the lowest mass objects had cooled sufficiently at their centers that any inbound dust collected at the center without melting. Some transitional objects would have a small molten core which cooled sufficiently quickly that its outer layers never melted in a hot core. The still more massive objects would have retained a molten core longer and developed density stratified layers of minerals from later generations of supernovae. Collisions of the mBD would have re-heated the cores and redistributed the stratified layers to form ores of the elements and primary compounds. The objects probably retained their extensive hydrogen envelopes, until interactions in the pre-solar accretion disc stripped away the hydrogen. Since the most massive mBD objects in this pre-solar processing would have been hotter if more massive, The HGD formation picture makes the qualitative prediction that the presently observable mean density should be highest for the most massive mBDs.

At the time of the Sun's formation, a million of these objects would have aggregated as the solar mass, leaving only nine planets in a pre-solar nebula with an accretion disc. Violent collisional events would have dumped most of the objects into the forming sun, ultimately stripping most bodies of their hydrogen and leaving the cores in the orbital plane.

In the pre-solar nebula, also called the accretion disc plane, a second period of structural modification took place. Collisions between pre-planetary bodies created meteoritic material, and re-heated and smeared the planetary interiors, mixing the metallic ores, and ejecting some of the bodies entirely. In this phase higher local densities and velocities would have stripped all bodies of their residual hydrogen outer atmospheres, and left nine planets, small meteoritic rocks, larger asteroid fragments, and unprocessed material in the distant Kuiper Belt and Oort cloud.

## 7. Surprises from the NASA's space missions STARDUST and DEEP IMPACT

An important confirmation of the 2-epoch planet formation scenario is seen in the quite unexpected results of the 2004 STARDUST space mission, in which dust grains released by Comet Temple I were collected in a jell that flew by the astronomical body (Brownlee, 2009). Two surprising aspects of the returned samples point to the same unexpected physics. The chemical composition of the samples, rich in Calcium Aluminum Inclusions, indicated that the chemistry of their formation occurred in a very high temperature environment, so they could not have formed where the comet has been stored for millenia in the outer solar system. A second surprise was that the captured cometary material was rich in molten chondrules that again indicated formation in a hot environment, not the outer Solar system where comets hibernate. Today planetary scientists ponder how the solar system turned itself inside-out, but in the primordial formation scenario, more probably these materials formed when the primordial planets were still hot before losing hydrogen, and before the Solar System formation had begun.

Another NASA space mission with surprising results was the Deep Impact mission to Comet Temple 1 to propel a 700 lb mass into the surface and examine the resulting dust cloud (A'Hearn, 2009). Instead of finding the rocks expected in the debris cloud, only a fine powder was found in the subsurface ejecta. This demonstrated that the ancient comet had collected dust throughout the history of the universe, but the dust grains had not collided and stuck together as predicted in the standard dust collision models. The powdery dust was so fine that its cloud saturated the detectors in the spacecraft.

## 8. Conclusions.

Recent Solar System spacecraft missions have produced results that have taken the community by surprise. Chemical composition and molten chondrules have forced the conclusion that the system has been turned inside out, with no previous evidence of this. And the existence of fine dust with no rocks in outer solar system bodies defies the theory that dust grains easily collide and stick together to form crystalline rocks. And the long-standing problem of how does the Earth come to have an iron core surmounted by an ocean of water has heretofore been unexplained.

While these sweeping statements describe reasonably the state of knowledge of solar system structure, they leave many unfinished details. The key point here is that planet formation was a 2-step process, with primordial rogue planets creating solid and semi-solid cores, and pre-solar disc evolution creating the presently observed structures near the sun (fig. 3 lower left). What has not been taken into account is the existence of not just the inner nine planets of the sun, but the many millions of primordial planets per star.

In this way, planetary formation does not need to begin with rocky particles colliding and sticking together on all mass scales from dust to moons.  The key ingredient is the existence of primordial gas rogue planets with extensive hydrogen atmospheres that collect interstellar material and stratify it by density at molten centers, so that when hydrogen-stripped in the pre-solar disc, their modified remnants supply all the ingredients of the observed solar system.